%% file: symmetry.tex
\documentstyle[12pt]{article}
\begin{document}
\baselineskip=5mm
\newcount\sectionnumber
\sectionnumber=0
\pagestyle{empty}
\begin{flushright}{ UTPT--96--08}
\end{flushright}

\vspace{8mm}
\begin{center}
{\bf {\huge The Spin--Symmetry of the Quark Model} }\\
\vspace{6mm}
Patrick J. O'Donnell, Q.P. Xu  \\
Physics Department,\\
University of Toronto,\\
Toronto, Ontario M5S 1A7, Canada,\\
\ \\
and \\
\ \\
Humphrey K.K. Tung\\
Department of Applied Physics\\
Hong Kong Polytechnic University\\
Hung Hom, Hong Kong.
\end{center}

\today

\vskip 20pt
\centerline{\bf Abstract}

Corrections  to  the  exact  heavy--quark  symmetry  results  are
expected to follow the $1/m_{Q}$ mass effect of the heavy--quark.
We show, by an  explicit  calculation,  that  there is  something
other than the mass effect that  suppresses  the  breaking of the
spin symmetry.

\newpage

\section{Introduction}

The  heavy--quark  symmetry,  which  appears in the  heavy--quark
limit,  gives  exact  results  for the  decays  of heavy  hadrons
\cite{heavy}.  Due to the heavy--quark  symmetry all form factors
in the  heavy--to--heavy  type of decays  such as $B  \rightarrow
D^{(\ast)} e {\bar \nu_e}$  ($D^{(\ast)}=D$  or $D^\ast$)  can be
related,  in  the  heavy--quark  limit,  to  a  single  universal
function   called  the  Isgur--Wise   function.  The  Isgur--Wise
function  is of  nonperturbative  origin  and has  been of  great
interest to both  theoretical and  experimental  studies.  In the
heavy--quark  symmetry limit, the decoupling of the  heavy--quark
spin with other light fields  leads to symmetry  relations  among
hadronic  matrix  elements.  The  corrections to the symmetry are
expected to follow the $1/m_{Q}$ mass effect of the heavy--quark.
It  has  been  unclear  how  well  these  would   extrapolate  to
heavy--to--light  quark  decays,  although  presumably  the charm
quark might still be heavy enough.

Many of these symmetries were anticipated in some versions of the
constituent quark model but there has not been an estimate of how
much of these were an  artifact  of the quark  model,  and of the
choice  of  wave   functions.  In  our  recent  paper  using  the
relativistic quark model \cite{xu}, we found that the breaking of
the spin symmetry  among  hadronic form factors is small even for
heavy--to--light   quark   decays.  In  this   paper,   we   show
explicitly,  using the same model, that there is something  other
than the mass effect that  suppresses  the  breaking  of the spin
symmetry.  In fact, the  quark  model  keeps  the  spin--symmetry
rules remarkably well for a wide range of masses.

We first recapitulate some aspects of spin symmetry for mesons in
the heavy--quark  limit.  For a pseudoscalar meson  $P(Q\bar{q})$
with heavy constituent  quark $Q$, the spin of $Q$ decouples from
all  other  light  fields  in  $P$  \cite{IW}.  We can  therefore
construct the spin operator  $S^{Z}_{Q}$ for $Q$ such that in the
heavy $m_{Q}$ limit
\begin{equation}
S^{Z}_{Q}|P(Q\bar{q})\rangle = \frac{1}{2}|V_{L}(Q\bar{q})\rangle
\,\,\, ,
\label{spin}
\end{equation}
where $V_{L}(Q\bar{q})$ is the longitudinal component of a vector
meson with the same quark  content as $P$.  In practice, the spin
symmetry in Eq.  (\ref{spin}) can be transformed  into identities
between the hadronic  matrix  elements, and thus some form factor
relations, for  $H\rightarrow P$ and $H\rightarrow  V_{L}$, where
$H(h\bar{q})$  is a pseudoscalar  meson with a heavy--quark  $h$.
Using             the             relation              \cite{IW}
$[S^{Z}_{Q},A^{0}+A^{3}]=(1/2)(V^{0}+V^{3})$   for  the  currents
$V_{\mu}=\bar{Q}\gamma_{\mu}h$                                and
$A_{\mu}=\bar{Q}\gamma_{\mu}\gamma_{5}h$    of   the   transition
$h\rightarrow Q$, it can be shown that Eq.  (\ref{spin}) leads to
the following  identity  between the hadronic matrix elements for
$H\rightarrow V_{L}$ and $H\rightarrow P$,
\begin{equation}
\langle V_{L}|A^{0}+A^{3}|H\rangle = \langle P|V^{0}+V^{3}|H\rangle
\,\,\, .
\label{matrix}
\end{equation}
In $|P\rangle$ and  $|V_{L}\rangle$  the spatial  momentum of the
quark $Q$ is defined in the  $z$--direction for the $Q$ spinor to
be an eigenstate of  $S^{Z}_{Q}$.  The spatial momenta of $P$ and
$V_{L}$ should  therefore  also be defined in the  $z$--direction
such that the  correction to the spin symmetry is of the order of
$\Lambda/m_{Q}$,  where $\Lambda$ is the internal energy scale of
$P$ and $V_{L}$.

\section{Kinematics}

In this paper, we  consider  the  breaking  of the spin  symmetry
coming from a finite quark mass $m_{Q}$ by directly  calculating,
in   particular,   the   hadronic    matrix   elements   in   Eq.
(\ref{matrix}).  We use the  relativistic  quark model formulated
in the infinite momentum frame (or equivalently, the light--front
quark  model)  \cite{xu,BSW,jaus,pat}.  We first define the ratio
of the matrix elements in Eq.  (\ref{matrix}) as

\begin{equation}
\rho (q^{2}) = \frac{\langle P(k')|V^{0}+V^{3}|H(p)\rangle}
                    {\langle V_{L}(k)|A^{0}+A^{3}|H(p)\rangle}
\,\,\, ,
\label{rho}
\end{equation}
so that $1-\rho$  represents the percentage  breaking of the spin
symmetry.  The ratio  $\rho$ is a function  of momentum  transfer
such  that  $\langle   V_{L}|A^{0}+A^{3}|H\rangle$  and  $\langle
P|V^{0}+V^{3}|H\rangle$  are evaluated at the same $q^{2}$.  This
allows us to evade the usual  problem of kinematic  discrepancies
when we come to  consider  a number of  different  final  states.
Since  the  spatial  momenta  of  $P$  and  $V_{L}$  are  in  the
$z$--direction,  we define the  function  $\rho$ in a frame where
the parametrization of the momenta in the $z$--direction is given
by

\begin{eqnarray}
p^{\mu} &=& (E_{H};0,0,p^{z}) \,\,\, , \nonumber\\
k^{\mu} &=& (E_{V};0,0,k^{z}) \,\,\, , \nonumber\\
k'^{\mu} &=& (E_{P};0,0,k'^{z}) \,\,\, .
\label{frame}
\end{eqnarray}

The vector and scalar  masses  $m_{V}$ and $m_{P}$ are  different
for finite  $m_{Q}$,  so $V_{L}$  and $P$ will not carry the same
momentum  even though the initial state $H$ has the same $p^{z}$.
We write the momenta  $k^{z}$ and  $k'^{z}$ in terms of the frame
parameter        $p^{z}$        through       the       condition
$q^{2}=(p-k)^{2}=(p-k')^{2}$.  The  general   parametrization  in
(\ref{frame})  includes  the  particular  case  of  the  infinite
momentum frame in which  $k^{z}=k'^{z}=p^{z}={\em P}$ where ${\em
P}\rightarrow\infty$ and $q^{2}=0$.  It is important to note that
the function  $\rho$, when  calculated  in the infinite  momentum
frame, is defined at  $q^{2}=0$  only.  This is also the point of
maximum  recoil,  which  is  usually  difficult  to  treat  in  a
non--relativistic  quark model, since a large amount of energy is
given to the  outgoing  particle.  In the  infinite  momentum  or
light--front  frame, we have the following  connection for $\rho$
to the form--factors defined in Ref.  \cite{xu};
\begin{equation}   
\rho^{IMF}  =  \frac{F_{1}^{H   \rightarrow
P}(0)}{A_{0}^{H   \rightarrow   V}(0)}   \,\,\,   .   
\label{IMF}
\end{equation}
However, here we shall calculate  $\rho^{IMF}$  directly from the
matrix element.

The  mass--shell  conditions  for  $p$,  $k$, and  $k'$  give the
following constraints on the momenta in (\ref{frame})
\begin{eqnarray}
\frac{m_{V}}{m_{P}}
\left( \frac{E_{P}+k'^{z}}{E_{V}+k^{z}} \right)
&=& \frac{w-\sqrt{w^{2}-1}}{w'-\sqrt{w'^{2}-1}}\,\,\, ,
\label{constraint}
\end{eqnarray}
where $w=p\cdot k/(m_{H}m_{V})$ and $w'=p\cdot  k'/(m_{H}m_{P})$.
The ratio $(E_{P}+k'^{z})/(E_{V}+k^{z})$ in (\ref{constraint}) is
therefore invariant for the frame defined in (\ref{frame}) and is
a     function     of    $q^{2}$     through     the     relation
$q^{2}=(p-k)^{2}=(p-k')^{2}$.

It can be shown  from the  covariant  expansion  of the  hadronic
matrix      elements       \cite{neubert}      that      $\langle
P(k')|V^{0}+V^{3}|H(p)\rangle   /(E_{P}+k'^{z})$   and   $\langle
V_{L}(k)|A^{0}+A^{3}|H(p)\rangle  /(E_{V}+k^{z})$  are  invariant
with  respect to the frame  defined in  (\ref{frame}).  Using the
kinematic  constraint  in  (\ref{constraint}),  it is easy to see
that the function  $\rho(q^{2})$  is an invariant  quantity.  The
matching of $p^{z}$ in $H\rightarrow  V_{L}$ and $H\rightarrow P$
of $\rho$ is the only choice that would lead to this  invariance.
In the definition of $\rho(q^{2})$,  there is an ambiguity coming
from the fact  that the  ranges  of  $q^{2}$  are  usually  quite
different in $H\rightarrow  V_{L}$ and $H\rightarrow P$.  We will
therefore  consider the value of $\rho$ at  $q^{2}=0$  only.  The
hadronic   matrix  elements  in  (\ref{rho})  can  be  calculated
reliably  using the  relativistic  quark model  formulated in the
infinite momentum frame.  So at $q^{2}=0$, we can write

\begin{equation}
\rho(0)=\rho^{IMF}
\,\,\, ,
\label{invariant}
\end{equation}
where  $\rho^{IMF}$ is calculated in the infinite momentum frame.

\section{Symmetry Breaking}

A brief  introduction  to the  relativistic  quark  model  in the
infinite momentum frame can be found in  Refs.\cite{xu,jaus,pat}.
In the relativistic quark model, the wave function for the ground
state meson $M(Q\bar{q})$ is given by
\begin{equation} 
|M({\bf k})\rangle = 
\sqrt{2}\int d{\bf p}_{Q} \sum_{\sigma\,\bar{\sigma}}
\Psi^{Jm_{J}}_{M,\,\sigma\bar{\sigma}}
|Q({\bf p}_{Q},\sigma)
\bar{q}({\bf k}-{\bf p}_{Q},\bar{\sigma})\rangle
\,\,\, ,
\end{equation}
where ${\bf k}={\em  P}\hat{{\bf z}}$ is the spatial  momentum of
the meson $M$,  ${\bf  p}_{Q}=({\bf  p}_{T},x{\em  P})$ and ${\bf
p}_{\bar{q}}={\bf k}-{\bf p}_{Q}=(-{\bf p}_{T},(1-x){\em P})$ are
those  of the  quarks  $Q$ and  $\bar{q}$,  respectively,  in the
infinite  momentum  frame.  Here,  $\Psi$  is the  momentum  wave
function for the  $Q\bar{q}$  bound state.  It has the  separable
form     into    the    spin     and     orbital     parts     as
$\,\Psi^{Jm_{J}}_{M,\,\sigma\bar{\sigma}}                       =
R^{Jm_{J}}_{M,\,\sigma\bar{\sigma}}\,\phi_{M}\,$,    where    the
expressions   for    $R^{Jm_{J}}_{M,\,\sigma\bar{\sigma}}$    and
$\phi_{M}$ can be found in Ref.  \cite{xu,pat}.

In  the  relativistic   quark  model,  it  has  been  shown  that
$\rho^{IMF}$  is a function of the mass  ratios $r$, $s$ and $l$,
so that $\,\rho(0)=\rho^{IMF}(r,s,l)\,$, where
\[
r=\frac{m_{Q}}{m_{h}}\,\,\, ,
s=\frac{m_{\bar{q}}}{m_{h}}\,\,\, ,
l=\frac{\Lambda}{m_{h}}\,\,\, .
\]
The parameter  $\Lambda$  determines the internal energy scale of
the  meson and  should be of the  order of  $\Lambda_{QCD}$.  The
dependence on $l$ appears only in the momentum  wave function and
actually  there  could be a  separate  $\Lambda$  for each of the
mesons  resulting in three further  parameters.  We take them all
to be equal here.  The kinematic region of interest for $r$, $s$,
and $l$ is such that $\,0 < l , r, s \leq 1\,$.  We find,
\begin{equation}
\rho^{IMF} = 
\frac{{\cal I}_{1}}{{\cal I}_{2}}\,\,\, ,
\label{rhoimf}
\end{equation}
where
\begin{equation}
{\cal I}_{1} =
\int^{1}_{0} dx \int^{\infty}_{0} dy \, y\,
\phi_{H}\,\phi_{P}\frac{\alpha_{0}(1,s)\alpha_{0}(r,s)+y^{2}\,}
{d_{0}(1,s)d_{0}(r,s) }
\end{equation}
and 
\begin{equation}
{\cal I}_{2} =
\int^{1}_{0} dx \int^{\infty}_{0} dy \, y\,
\phi_{H}\,\phi_{V}\frac{
\,\alpha_{0}(1,s)\alpha_{1}(r,s)\alpha_{2}(r,s)+y^{2}
\left[\,\alpha_{1}(r,s)-\alpha_{2}(r,s)+\alpha_{0}(1,s)\,\right]
\,}{d_{0}(1,s)d_{1}(r,s)d_{2}(r,s)} \, ,
\end{equation}
with the definitions
\begin{eqnarray}
\alpha_{0}(r,s) &=& xs+(1-x)r \,\, , \,\,
\alpha_{1}(r,s) = r+xM_{0}(r,s) \,\, , \,\,
\alpha_{2}(r,s) = s+(1-x)M_{0}(r,s)\,\, , \, \,\nonumber\\
M_{0}(r,s) &=& \sqrt{\frac{r^{2}+y^{2}}{x}
                 + \frac{s^{2}+y^{2}}{1-x}}\,\,\, ,\nonumber\\
d_{0}(r,s) &=& \sqrt{\alpha^{2}_{0}(r,s)+y^{2}}\,\, , \,\,
d_{1}(r,s) = \sqrt{\alpha^{2}_{1}(r,s)+y^{2}}\,\, , \,\,
d_{2}(r,s) = \sqrt{\alpha^{2}_{2}(r,s)+y^{2}}\,\, .\nonumber 
\end{eqnarray}

We  may  write  the  orbital  wave   functions   $\phi_{H}$   and
$\phi_{P,V}$ in terms of a Gaussian  function $\phi (r,s,l)$ such
that $\phi_{H}=\phi (1,s,l)$ and $\phi_{P,V}=\phi  (r,s,l)$.  The
expression     for     $\phi     (r,s,l)$     is     given     by
\cite{xu,jaus,pat,gauss}
\begin{equation}
\phi (r,s,l)=N\sqrt{\frac{dz}{dx}}
exp\left(\, -\frac{1}{2}(y^{2}+z^{2})/l^2\,\right)
\,\,\, ,
\label{gaussfunction}
\end{equation}
where $N$ is a  normalization  factor  that is  canceled  out in
$\rho^{IMF}$, and
\[
z=(x-\frac{1}{2})M_{0}(r,s)-\frac{(r^{2}-s^{2})}{2M_{0}(r,s)}
\,\,\, .
\]
In Ref.  \cite{xu},  it has been  pointed  out that  the  scaling
behavior of the meson decay constant  $f_{M}$ in the heavy--quark
limit  imposes a constraint  on the orbital  wave  function.  The
Gaussian  wave  function  in  (\ref{gaussfunction})  is shown  to
satisfy the scaling law  $1/\sqrt{m_{Q}}$ of $f_{M}$ in the heavy
$m_{Q}$ limit.

In the numerical  analysis of $\rho(0)$, it is  convenient to set
$s=l$ and vary $s$ and $r$ within the kinematic  region of $\,0 <
s  ,  r  \leq  1\,$.  The  spectator   quark  $\bar{q}$  is  thus
considered to be a light quark with  $m_{\bar{q}}  = \Lambda \sim
\Lambda_{\rm  QCD}$.  In case  of a  heavy--quark  decaying  to a
heavy or to a light quark, the corresponding  regions for $s$ and
$r$ are such that $s$ is small and $r$  varies  between  1 and 0.
By comparison, for a light quark decay to another light quark, we
look at the region  where $s$ and $r$ could both be close to 1.  When
$s=r=1$,   the   symmetry    breaking   is   calculated   to   be
$1-\rho(0)=-0.33$,   using  the   Gaussian   wave   function   in
(\ref{gaussfunction}).  Thus the  heavy--quark  symmetry does not
apply in that case, as expected.

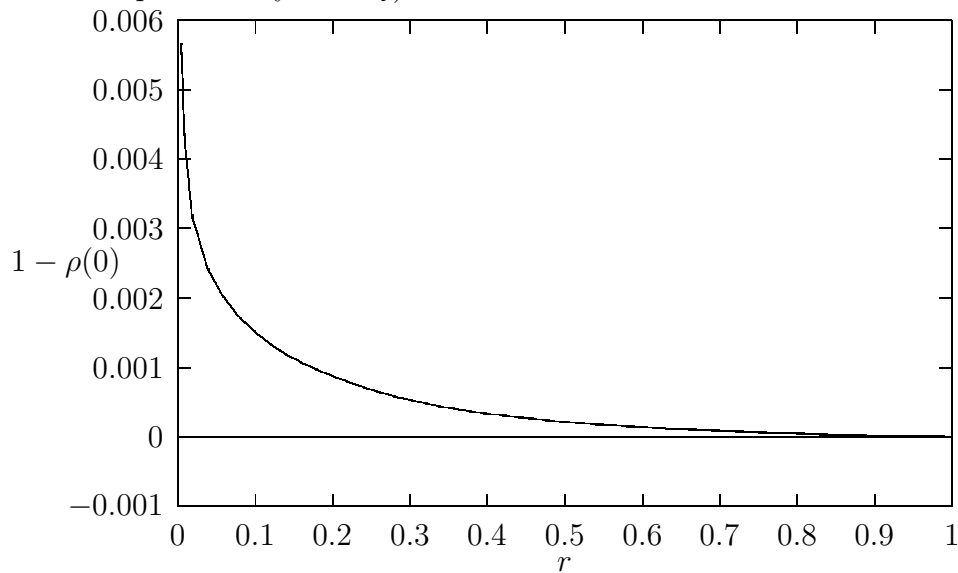
\begin{figure}
\caption{The plot of $1-\rho(0)$ using the Gaussian orbital 
wave function. The plot is for heavy--to--heavy and 
heavy--to--light decays with a quark mass ratio $s=l=0.001$ and where 
$r$ varies within the range of $0 < r \leq 1$.  
($r$ refers to the ratio
$r=m_Q/m_h$ in the quark decay $h \rightarrow Q$).
\label{fig1}}
\input{fig1.tex}
\end{figure}

In Fig.  (\ref{fig1}), we show the plot of $1-\rho(0)$ for a {\it
very}  heavy--quark  decaying  to a heavy or to a light  quark so
that the the mass  ratios $s$ and $l$ are very  small  (here they
are taken to be $s=l=0.001$).  The variation of $r$ is within the
range  $0 < r \leq  1$.  From  the  Gaussian  form  of  the  wave
function  in  Eq.  (\ref{gaussfunction})   one  expects  $\langle
y^{2}\rangle=l^{2}$  and from the expression for  $\rho^{IMF}$ in
Eq.  (\ref{rhoimf}),  we  expect  $1-\rho(0)\rightarrow  0$ since
$\langle  y^{2}\rangle\rightarrow  0$ for a very  heavy  decaying
quark.

In Fig.  (\ref{fig1}),  we show that the breaking  $1-\rho(0)$ is
less than 1\% in the  heavy--quark  limit (or  small  $s=0.001$).
From the figure,  the mass effect of $m_{Q}$ can be seen  clearly
as the breaking of the spin  symmetry  gradually  increases  from
large $r$, or  heavy--to--heavy  decays, towards the smaller $r$,
or  heavy--to--light  region.  However,  even at very small  $r$,
corresponding   to  a  decay   $b\rightarrow   s$  ($r=0.1$)   or
$b\rightarrow  u$ ($r=0.06$), it is remarkable  that the symmetry
breaking is less than $0.6\%$.

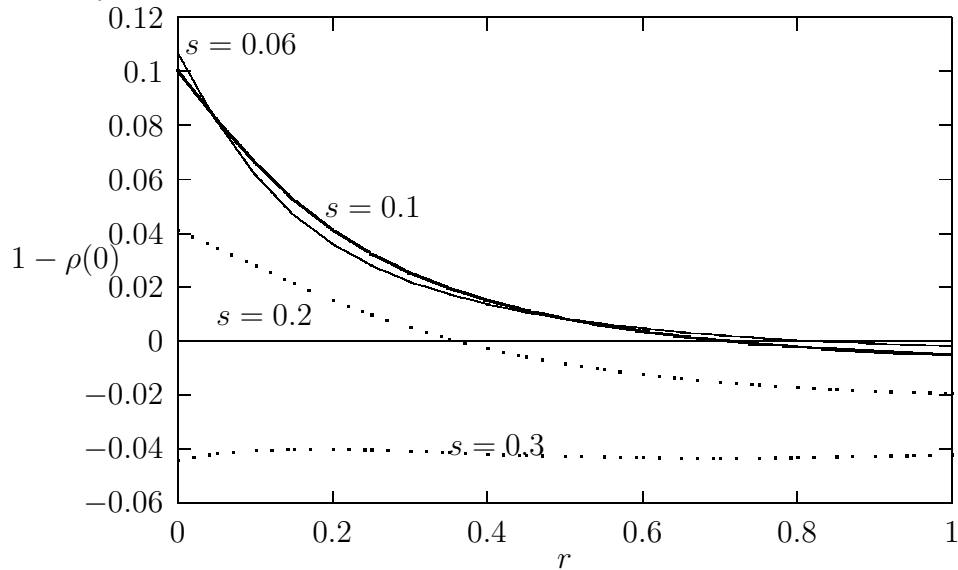
\begin{figure}  
\caption{ 
The   plot   of    $1-\rho(0)$    with   physical   mass   ratios
$s=m_{\bar{u}}/m_{b}=0.06$,   $s=m_{\bar{s}}/m_{b}=0.1$,     
$s=m_{\bar{u}}/m_{c}=0.2$ and
$s=m_{\bar{s}}/m_{c}=0.3$,  for heavy $b$
and $c$ quark  decays.  
\label{fig2}}
\input{fig2.tex}
\end{figure}

In Fig.  (\ref{fig2}),  we show the plot of $1-\rho(0)$  with the
physical  spectator quark mass ratios for heavy--quark  decays as
$s=m_{\bar{u}}/m_{b}=0.06$,            $s=m_{\bar{s}}/m_{b}=0.1$,
$s=m_{\bar{u}}/m_{c}=0.2$, and $s=m_{\bar{s}}/m_{c}=0.3$.  In the
figure,  the  breaking of the spin  symmetry  is shown to be less
than  about  10\% for  heavy  $b$ and $c$  quark  decays.  In the
strict heavy mass limit of  $s\rightarrow0$  as indicated in Fig.
(\ref{fig1})  , the mass effect of $m_{Q}$ is clearly seen as the
suppression  of  the  symmetry   breaking  with  increasing  $r$.
Notice,  however,  that for finite  spectator  mass  ratios,  the
function  $1-\rho(0)$  passes  through  zero  at  a  recoil  mass
$r=m_{Q}/m_{h}$  below its heaviest limit $r=1$.  

In the case of $b$ decays and the decay of charm to  non--strange
quarks, the mass effect of $m_Q$is still seen in the region where
$1 -  \rho(0)$  is  positive  for a large  range  of $r$.  In the
region where $1 - \rho(0)$ is negative, the $m_Q$  suppression no
longer  follows  as the  size  of $1 -  \rho(0)$  increases  with
$r=m_Q/m_h$.  This  suggests  that  there are  kinematic  factors
other than the mass effect of $m_{Q}$ that govern the size of the
symmetry breaking $1-\rho(0)$, when the decaying quark has finite
mass.  As shown in the  figure,  the  zero of  $1-\rho(0)$  is at
smaller   $r$  when  the  ratio  $s$  is  larger.  The  zero  for
$s\rightarrow  0$ appears at  $r\rightarrow 1$ and decreases with
increasing $s$.  For $s \geq 0.251$,  $1-\rho(0)$ is negative for
all $r$.  The mass effect of $m_{Q}$ is therefore less pronounced
as  $s$  gets  larger  and  the  kinematic  effects  dominate.  A
different  type of  behavior  enters for the decay  involving  a
charm quark and a strange  spectator where the deviation from the
symmetry limit is rather constant and about $4\%$.

\begin{figure}
\caption{The corresponding values of $s$ and $r$ for which 
$1-\rho(0)=0$ and $1-\rho(0)=\pm 0.01$. 
\label{fig3}}
\input{fig3.tex}
\end{figure}
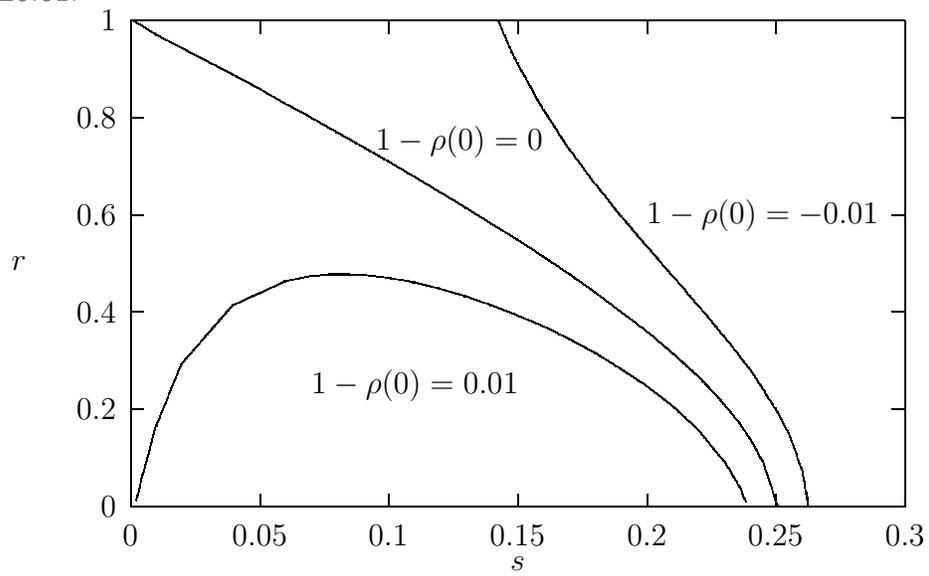

In fig.  (\ref{fig3}),  we show the  corresponding  values of $s$
and $r$ for which $1-\rho(0)=0$.  The zero is shown to lie within
the region  where $s$ is small ($s \leq 0.251$) and the  decaying
quark is heavy.  As shown in the figure, the zero of  $1-\rho(0)$
appears at smaller $r$ as the mass of the decaying  quark becomes
less heavy  (larger $s$).  Also shown in the figure are the plots
for  which the  spin--symmetry  breaking  is about  10\%  that is
$|1-\rho(0)|=0.01$.  It can been seen that a large portion of the
possible  phase  space of $r$ and $s$ is within the region  where
$|1-\rho(0)| \leq 0.01$.

We have  also  obtained  a  similar  result  using  the  harmonic
oscillator  wave  function  \cite{BSW}  instead  of the  Gaussian
function.  The  result  is   therefore   not  an  artifact  of  a
particular  momentum wave  function.  The quantity  $\rho(0)$ has
the following physical meaning:
\begin{equation}       
|\rho(0)|^{2}=
\frac{(m^{2}_{H}-m^{2}_{V})^{3}}{(m^{2}_{H}-m^{2}_{P})^{3}}
\frac{ d\Gamma  (H\rightarrow  Pl\bar{\nu})/dq^{2}|_{q^{2}=0}  }{
d\Gamma (H\rightarrow  V_{L}l\bar{\nu})/dq^{2}|_{q^{2}=0} }\,\,\,.  
\end{equation}  
This allows a test of these results to be made by considering the
$q^{2}$     spectrum     for     the     semileptonic      decays
$H(h\bar{q})\rightarrow    P,V_{L}(Q\bar{q})$.   The    size   of
$\rho(0)$  for  particular  values  of $r$  and  $s$  can  now be
measured.  Repeating  this for the different  semileptonic  decay
channels of $H$, the dependence of $\rho(0)$ with $r$ and $s$ can
also be determined.

\centerline{ {\bf ACKNOWLEDGMENTS}}

This work was supported by the Natural Sciences and Engineering
Council of Canada.

\newpage

\end{document}

%% file: fig1.tex
\setlength{\unitlength}{0.240900pt}
\ifx\plotpoint\undefined\newsavebox{\plotpoint}\fi
\sbox{\plotpoint}{\rule[-0.150pt]{0.300pt}{0.300pt}}%
\begin{picture}(1500,900)(0,0)
\font\gnuplot=cmr10 at 10pt
\gnuplot
\sbox{\plotpoint}{\rule[-0.150pt]{0.300pt}{0.300pt}}%
\put(220.0,222.0){\rule[-0.150pt]{292.934pt}{0.300pt}}
\put(220.0,113.0){\rule[-0.150pt]{0.300pt}{184.048pt}}
\put(220.0,113.0){\rule[-0.150pt]{4.818pt}{0.300pt}}
\put(198,113){\makebox(0,0)[r]{$-0.001$}}
\put(1416.0,113.0){\rule[-0.150pt]{4.818pt}{0.300pt}}
\put(220.0,222.0){\rule[-0.150pt]{4.818pt}{0.300pt}}
\put(198,222){\makebox(0,0)[r]{$0$}}
\put(1416.0,222.0){\rule[-0.150pt]{4.818pt}{0.300pt}}
\put(220.0,331.0){\rule[-0.150pt]{4.818pt}{0.300pt}}
\put(198,331){\makebox(0,0)[r]{$0.001$}}
\put(1416.0,331.0){\rule[-0.150pt]{4.818pt}{0.300pt}}
\put(220.0,440.0){\rule[-0.150pt]{4.818pt}{0.300pt}}
\put(198,440){\makebox(0,0)[r]{$0.002$}}
\put(1416.0,440.0){\rule[-0.150pt]{4.818pt}{0.300pt}}
\put(220.0,550.0){\rule[-0.150pt]{4.818pt}{0.300pt}}
\put(198,550){\makebox(0,0)[r]{$0.003$}}
\put(1416.0,550.0){\rule[-0.150pt]{4.818pt}{0.300pt}}
\put(220.0,659.0){\rule[-0.150pt]{4.818pt}{0.300pt}}
\put(198,659){\makebox(0,0)[r]{$0.004$}}
\put(1416.0,659.0){\rule[-0.150pt]{4.818pt}{0.300pt}}
\put(220.0,768.0){\rule[-0.150pt]{4.818pt}{0.300pt}}
\put(198,768){\makebox(0,0)[r]{$0.005$}}
\put(1416.0,768.0){\rule[-0.150pt]{4.818pt}{0.300pt}}
\put(220.0,877.0){\rule[-0.150pt]{4.818pt}{0.300pt}}
\put(198,877){\makebox(0,0)[r]{$0.006$}}
\put(1416.0,877.0){\rule[-0.150pt]{4.818pt}{0.300pt}}
\put(220.0,113.0){\rule[-0.150pt]{0.300pt}{4.818pt}}
\put(220,68){\makebox(0,0){$0$}}
\put(220.0,857.0){\rule[-0.150pt]{0.300pt}{4.818pt}}
\put(342.0,113.0){\rule[-0.150pt]{0.300pt}{4.818pt}}
\put(342,68){\makebox(0,0){$0.1$}}
\put(342.0,857.0){\rule[-0.150pt]{0.300pt}{4.818pt}}
\put(463.0,113.0){\rule[-0.150pt]{0.300pt}{4.818pt}}
\put(463,68){\makebox(0,0){$0.2$}}
\put(463.0,857.0){\rule[-0.150pt]{0.300pt}{4.818pt}}
\put(585.0,113.0){\rule[-0.150pt]{0.300pt}{4.818pt}}
\put(585,68){\makebox(0,0){$0.3$}}
\put(585.0,857.0){\rule[-0.150pt]{0.300pt}{4.818pt}}
\put(706.0,113.0){\rule[-0.150pt]{0.300pt}{4.818pt}}
\put(706,68){\makebox(0,0){$0.4$}}
\put(706.0,857.0){\rule[-0.150pt]{0.300pt}{4.818pt}}
\put(828.0,113.0){\rule[-0.150pt]{0.300pt}{4.818pt}}
\put(828,68){\makebox(0,0){$0.5$}}
\put(828.0,857.0){\rule[-0.150pt]{0.300pt}{4.818pt}}
\put(950.0,113.0){\rule[-0.150pt]{0.300pt}{4.818pt}}
\put(950,68){\makebox(0,0){$0.6$}}
\put(950.0,857.0){\rule[-0.150pt]{0.300pt}{4.818pt}}
\put(1071.0,113.0){\rule[-0.150pt]{0.300pt}{4.818pt}}
\put(1071,68){\makebox(0,0){$0.7$}}
\put(1071.0,857.0){\rule[-0.150pt]{0.300pt}{4.818pt}}
\put(1193.0,113.0){\rule[-0.150pt]{0.300pt}{4.818pt}}
\put(1193,68){\makebox(0,0){$0.8$}}
\put(1193.0,857.0){\rule[-0.150pt]{0.300pt}{4.818pt}}
\put(1314.0,113.0){\rule[-0.150pt]{0.300pt}{4.818pt}}
\put(1314,68){\makebox(0,0){$0.9$}}
\put(1314.0,857.0){\rule[-0.150pt]{0.300pt}{4.818pt}}
\put(1436.0,113.0){\rule[-0.150pt]{0.300pt}{4.818pt}}
\put(1436,68){\makebox(0,0){$1$}}
\put(1436.0,857.0){\rule[-0.150pt]{0.300pt}{4.818pt}}
\put(220.0,113.0){\rule[-0.150pt]{292.934pt}{0.300pt}}
\put(1436.0,113.0){\rule[-0.150pt]{0.300pt}{184.048pt}}
\put(220.0,877.0){\rule[-0.150pt]{292.934pt}{0.300pt}}
\put(45,495){\makebox(0,0){$1-\rho(0)$}}
\put(828,23){\makebox(0,0){$r$}}
\put(220.0,113.0){\rule[-0.150pt]{0.300pt}{184.048pt}}
\put(226,840){\usebox{\plotpoint}}
\multiput(226.38,806.89)(0.475,-13.877){10}{\rule{0.115pt}{7.975pt}}
\multiput(225.38,823.45)(6.000,-141.447){2}{\rule{0.300pt}{3.988pt}}
\multiput(232.38,669.65)(0.489,-4.936){22}{\rule{0.118pt}{2.975pt}}
\multiput(231.38,675.83)(12.000,-109.825){2}{\rule{0.300pt}{1.488pt}}
\multiput(244.38,561.60)(0.495,-1.651){48}{\rule{0.119pt}{1.059pt}}
\multiput(243.38,563.80)(25.000,-79.802){2}{\rule{0.300pt}{0.530pt}}
\multiput(269.38,481.51)(0.495,-0.878){46}{\rule{0.119pt}{0.600pt}}
\multiput(268.38,482.75)(24.000,-40.755){2}{\rule{0.300pt}{0.300pt}}
\multiput(293.38,440.08)(0.495,-0.646){46}{\rule{0.119pt}{0.463pt}}
\multiput(292.38,441.04)(24.000,-30.040){2}{\rule{0.300pt}{0.231pt}}
\multiput(317.00,410.13)(0.520,-0.495){46}{\rule{0.388pt}{0.119pt}}
\multiput(317.00,410.38)(24.196,-24.000){2}{\rule{0.194pt}{0.300pt}}
\multiput(342.00,386.13)(0.632,-0.493){36}{\rule{0.454pt}{0.119pt}}
\multiput(342.00,386.38)(23.058,-19.000){2}{\rule{0.227pt}{0.300pt}}
\multiput(366.00,367.13)(0.752,-0.492){30}{\rule{0.525pt}{0.118pt}}
\multiput(366.00,367.38)(22.910,-16.000){2}{\rule{0.263pt}{0.300pt}}
\multiput(390.00,351.13)(0.969,-0.490){24}{\rule{0.652pt}{0.118pt}}
\multiput(390.00,351.38)(23.647,-13.000){2}{\rule{0.326pt}{0.300pt}}
\multiput(415.00,338.13)(1.104,-0.488){20}{\rule{0.730pt}{0.117pt}}
\multiput(415.00,338.38)(22.486,-11.000){2}{\rule{0.365pt}{0.300pt}}
\multiput(439.00,327.13)(1.218,-0.486){18}{\rule{0.795pt}{0.117pt}}
\multiput(439.00,327.38)(22.350,-10.000){2}{\rule{0.398pt}{0.300pt}}
\multiput(463.00,317.14)(1.415,-0.485){16}{\rule{0.908pt}{0.117pt}}
\multiput(463.00,317.38)(23.115,-9.000){2}{\rule{0.454pt}{0.300pt}}
\multiput(488.00,308.14)(1.358,-0.485){16}{\rule{0.875pt}{0.117pt}}
\multiput(488.00,308.38)(22.184,-9.000){2}{\rule{0.438pt}{0.300pt}}
\multiput(512.00,299.14)(1.765,-0.480){12}{\rule{1.104pt}{0.116pt}}
\multiput(512.00,299.38)(21.709,-7.000){2}{\rule{0.552pt}{0.300pt}}
\multiput(536.00,292.14)(1.765,-0.480){12}{\rule{1.104pt}{0.116pt}}
\multiput(536.00,292.38)(21.709,-7.000){2}{\rule{0.552pt}{0.300pt}}
\multiput(560.00,285.14)(2.167,-0.475){10}{\rule{1.325pt}{0.115pt}}
\multiput(560.00,285.38)(22.250,-6.000){2}{\rule{0.663pt}{0.300pt}}
\multiput(585.00,279.14)(2.530,-0.469){8}{\rule{1.515pt}{0.113pt}}
\multiput(585.00,279.38)(20.856,-5.000){2}{\rule{0.758pt}{0.300pt}}
\multiput(609.00,274.14)(2.530,-0.469){8}{\rule{1.515pt}{0.113pt}}
\multiput(609.00,274.38)(20.856,-5.000){2}{\rule{0.758pt}{0.300pt}}
\multiput(633.00,269.15)(3.381,-0.459){6}{\rule{1.950pt}{0.111pt}}
\multiput(633.00,269.38)(20.953,-4.000){2}{\rule{0.975pt}{0.300pt}}
\multiput(658.00,265.15)(3.243,-0.459){6}{\rule{1.875pt}{0.111pt}}
\multiput(658.00,265.38)(20.108,-4.000){2}{\rule{0.938pt}{0.300pt}}
\multiput(682.00,261.16)(4.575,-0.439){4}{\rule{2.475pt}{0.106pt}}
\multiput(682.00,261.38)(18.863,-3.000){2}{\rule{1.238pt}{0.300pt}}
\multiput(706.00,258.16)(4.770,-0.439){4}{\rule{2.575pt}{0.106pt}}
\multiput(706.00,258.38)(19.655,-3.000){2}{\rule{1.288pt}{0.300pt}}
\multiput(731.00,255.16)(4.575,-0.439){4}{\rule{2.475pt}{0.106pt}}
\multiput(731.00,255.38)(18.863,-3.000){2}{\rule{1.238pt}{0.300pt}}
\multiput(755.00,252.19)(8.745,-0.377){2}{\rule{3.675pt}{0.091pt}}
\multiput(755.00,252.38)(16.372,-2.000){2}{\rule{1.838pt}{0.300pt}}
\multiput(779.00,250.16)(4.770,-0.439){4}{\rule{2.575pt}{0.106pt}}
\multiput(779.00,250.38)(19.655,-3.000){2}{\rule{1.288pt}{0.300pt}}
\multiput(804.00,247.19)(8.745,-0.377){2}{\rule{3.675pt}{0.091pt}}
\multiput(804.00,247.38)(16.372,-2.000){2}{\rule{1.838pt}{0.300pt}}
\multiput(828.00,245.19)(8.745,-0.377){2}{\rule{3.675pt}{0.091pt}}
\multiput(828.00,245.38)(16.372,-2.000){2}{\rule{1.838pt}{0.300pt}}
\multiput(852.00,243.19)(9.122,-0.377){2}{\rule{3.825pt}{0.091pt}}
\multiput(852.00,243.38)(17.061,-2.000){2}{\rule{1.913pt}{0.300pt}}
\put(877,240.88){\rule{5.782pt}{0.300pt}}
\multiput(877.00,241.38)(12.000,-1.000){2}{\rule{2.891pt}{0.300pt}}
\multiput(901.00,240.19)(8.745,-0.377){2}{\rule{3.675pt}{0.091pt}}
\multiput(901.00,240.38)(16.372,-2.000){2}{\rule{1.838pt}{0.300pt}}
\put(925,237.88){\rule{6.023pt}{0.300pt}}
\multiput(925.00,238.38)(12.500,-1.000){2}{\rule{3.011pt}{0.300pt}}
\multiput(950.00,237.19)(8.745,-0.377){2}{\rule{3.675pt}{0.091pt}}
\multiput(950.00,237.38)(16.372,-2.000){2}{\rule{1.838pt}{0.300pt}}
\put(974,234.88){\rule{5.782pt}{0.300pt}}
\multiput(974.00,235.38)(12.000,-1.000){2}{\rule{2.891pt}{0.300pt}}
\put(998,233.88){\rule{6.023pt}{0.300pt}}
\multiput(998.00,234.38)(12.500,-1.000){2}{\rule{3.011pt}{0.300pt}}
\put(1023,232.88){\rule{5.782pt}{0.300pt}}
\multiput(1023.00,233.38)(12.000,-1.000){2}{\rule{2.891pt}{0.300pt}}
\put(1047,231.88){\rule{5.782pt}{0.300pt}}
\multiput(1047.00,232.38)(12.000,-1.000){2}{\rule{2.891pt}{0.300pt}}
\put(1071,230.88){\rule{6.023pt}{0.300pt}}
\multiput(1071.00,231.38)(12.500,-1.000){2}{\rule{3.011pt}{0.300pt}}
\put(1096,229.88){\rule{5.782pt}{0.300pt}}
\multiput(1096.00,230.38)(12.000,-1.000){2}{\rule{2.891pt}{0.300pt}}
\put(1120,228.88){\rule{5.782pt}{0.300pt}}
\multiput(1120.00,229.38)(12.000,-1.000){2}{\rule{2.891pt}{0.300pt}}
\put(1144,227.88){\rule{5.782pt}{0.300pt}}
\multiput(1144.00,228.38)(12.000,-1.000){2}{\rule{2.891pt}{0.300pt}}
\put(1193,226.88){\rule{5.782pt}{0.300pt}}
\multiput(1193.00,227.38)(12.000,-1.000){2}{\rule{2.891pt}{0.300pt}}
\put(1217,225.88){\rule{5.782pt}{0.300pt}}
\multiput(1217.00,226.38)(12.000,-1.000){2}{\rule{2.891pt}{0.300pt}}
\put(1168.0,228.0){\rule[-0.150pt]{6.022pt}{0.300pt}}
\put(1266,224.88){\rule{5.782pt}{0.300pt}}
\multiput(1266.00,225.38)(12.000,-1.000){2}{\rule{2.891pt}{0.300pt}}
\put(1290,223.88){\rule{5.782pt}{0.300pt}}
\multiput(1290.00,224.38)(12.000,-1.000){2}{\rule{2.891pt}{0.300pt}}
\put(1241.0,226.0){\rule[-0.150pt]{6.022pt}{0.300pt}}
\put(1339,222.88){\rule{5.782pt}{0.300pt}}
\multiput(1339.00,223.38)(12.000,-1.000){2}{\rule{2.891pt}{0.300pt}}
\put(1314.0,224.0){\rule[-0.150pt]{6.022pt}{0.300pt}}
\put(1412,221.88){\rule{5.782pt}{0.300pt}}
\multiput(1412.00,222.38)(12.000,-1.000){2}{\rule{2.891pt}{0.300pt}}
\put(1363.0,223.0){\rule[-0.150pt]{11.804pt}{0.300pt}}
\end{picture}

%% file: fig2.tex
\setlength{\unitlength}{0.240900pt}
\ifx\plotpoint\undefined\newsavebox{\plotpoint}\fi
\sbox{\plotpoint}{\rule[-0.150pt]{0.300pt}{0.300pt}}%
\begin{picture}(1500,900)(0,0)
\font\gnuplot=cmr10 at 10pt
\gnuplot
\sbox{\plotpoint}{\rule[-0.150pt]{0.300pt}{0.300pt}}%
\put(220.0,368.0){\rule[-0.150pt]{292.934pt}{0.300pt}}
\put(220.0,113.0){\rule[-0.150pt]{0.300pt}{184.048pt}}
\put(220.0,113.0){\rule[-0.150pt]{4.818pt}{0.300pt}}
\put(198,113){\makebox(0,0)[r]{$-0.06$}}
\put(1416.0,113.0){\rule[-0.150pt]{4.818pt}{0.300pt}}
\put(220.0,198.0){\rule[-0.150pt]{4.818pt}{0.300pt}}
\put(198,198){\makebox(0,0)[r]{$-0.04$}}
\put(1416.0,198.0){\rule[-0.150pt]{4.818pt}{0.300pt}}
\put(220.0,283.0){\rule[-0.150pt]{4.818pt}{0.300pt}}
\put(198,283){\makebox(0,0)[r]{$-0.02$}}
\put(1416.0,283.0){\rule[-0.150pt]{4.818pt}{0.300pt}}
\put(220.0,368.0){\rule[-0.150pt]{4.818pt}{0.300pt}}
\put(198,368){\makebox(0,0)[r]{$0$}}
\put(1416.0,368.0){\rule[-0.150pt]{4.818pt}{0.300pt}}
\put(220.0,453.0){\rule[-0.150pt]{4.818pt}{0.300pt}}
\put(198,453){\makebox(0,0)[r]{$0.02$}}
\put(1416.0,453.0){\rule[-0.150pt]{4.818pt}{0.300pt}}
\put(220.0,537.0){\rule[-0.150pt]{4.818pt}{0.300pt}}
\put(198,537){\makebox(0,0)[r]{$0.04$}}
\put(1416.0,537.0){\rule[-0.150pt]{4.818pt}{0.300pt}}
\put(220.0,622.0){\rule[-0.150pt]{4.818pt}{0.300pt}}
\put(198,622){\makebox(0,0)[r]{$0.06$}}
\put(1416.0,622.0){\rule[-0.150pt]{4.818pt}{0.300pt}}
\put(220.0,707.0){\rule[-0.150pt]{4.818pt}{0.300pt}}
\put(198,707){\makebox(0,0)[r]{$0.08$}}
\put(1416.0,707.0){\rule[-0.150pt]{4.818pt}{0.300pt}}
\put(220.0,792.0){\rule[-0.150pt]{4.818pt}{0.300pt}}
\put(198,792){\makebox(0,0)[r]{$0.1$}}
\put(1416.0,792.0){\rule[-0.150pt]{4.818pt}{0.300pt}}
\put(220.0,877.0){\rule[-0.150pt]{4.818pt}{0.300pt}}
\put(198,877){\makebox(0,0)[r]{$0.12$}}
\put(1416.0,877.0){\rule[-0.150pt]{4.818pt}{0.300pt}}
\put(220.0,113.0){\rule[-0.150pt]{0.300pt}{4.818pt}}
\put(220,68){\makebox(0,0){$0$}}
\put(220.0,857.0){\rule[-0.150pt]{0.300pt}{4.818pt}}
\put(463.0,113.0){\rule[-0.150pt]{0.300pt}{4.818pt}}
\put(463,68){\makebox(0,0){$0.2$}}
\put(463.0,857.0){\rule[-0.150pt]{0.300pt}{4.818pt}}
\put(706.0,113.0){\rule[-0.150pt]{0.300pt}{4.818pt}}
\put(706,68){\makebox(0,0){$0.4$}}
\put(706.0,857.0){\rule[-0.150pt]{0.300pt}{4.818pt}}
\put(950.0,113.0){\rule[-0.150pt]{0.300pt}{4.818pt}}
\put(950,68){\makebox(0,0){$0.6$}}
\put(950.0,857.0){\rule[-0.150pt]{0.300pt}{4.818pt}}
\put(1193.0,113.0){\rule[-0.150pt]{0.300pt}{4.818pt}}
\put(1193,68){\makebox(0,0){$0.8$}}
\put(1193.0,857.0){\rule[-0.150pt]{0.300pt}{4.818pt}}
\put(1436.0,113.0){\rule[-0.150pt]{0.300pt}{4.818pt}}
\put(1436,68){\makebox(0,0){$1$}}
\put(1436.0,857.0){\rule[-0.150pt]{0.300pt}{4.818pt}}
\put(220.0,113.0){\rule[-0.150pt]{292.934pt}{0.300pt}}
\put(1436.0,113.0){\rule[-0.150pt]{0.300pt}{184.048pt}}
\put(220.0,877.0){\rule[-0.150pt]{292.934pt}{0.300pt}}
\put(45,495){\makebox(0,0){$1-\rho(0)$}}
\put(828,23){\makebox(0,0){$r$}}
\put(232,835){\makebox(0,0)[l]{$s=0.06$}}
\put(451,580){\makebox(0,0)[l]{$s=0.1$}}
\put(281,410){\makebox(0,0)[l]{$s=0.2$}}
\put(646,206){\makebox(0,0)[l]{$s=0.3$}}
\put(220.0,113.0){\rule[-0.150pt]{0.300pt}{184.048pt}}
\put(220,822){\usebox{\plotpoint}}
\multiput(220.37,819.50)(0.498,-0.878){120}{\rule{0.120pt}{0.601pt}}
\multiput(219.38,820.75)(61.000,-105.752){2}{\rule{0.300pt}{0.301pt}}
\multiput(281.37,712.95)(0.498,-0.697){120}{\rule{0.120pt}{0.493pt}}
\multiput(280.38,713.98)(61.000,-83.977){2}{\rule{0.300pt}{0.247pt}}
\multiput(342.37,628.38)(0.498,-0.525){118}{\rule{0.120pt}{0.390pt}}
\multiput(341.38,629.19)(60.000,-62.191){2}{\rule{0.300pt}{0.195pt}}
\multiput(402.00,566.13)(0.663,-0.497){90}{\rule{0.473pt}{0.120pt}}
\multiput(402.00,566.38)(60.019,-46.000){2}{\rule{0.236pt}{0.300pt}}
\multiput(463.00,520.13)(0.899,-0.496){66}{\rule{0.613pt}{0.120pt}}
\multiput(463.00,520.38)(59.727,-34.000){2}{\rule{0.307pt}{0.300pt}}
\multiput(524.00,486.13)(1.179,-0.495){50}{\rule{0.779pt}{0.119pt}}
\multiput(524.00,486.38)(59.383,-26.000){2}{\rule{0.389pt}{0.300pt}}
\multiput(585.00,460.13)(1.620,-0.493){36}{\rule{1.038pt}{0.119pt}}
\multiput(585.00,460.38)(58.845,-19.000){2}{\rule{0.519pt}{0.300pt}}
\multiput(646.00,441.13)(1.897,-0.492){30}{\rule{1.200pt}{0.118pt}}
\multiput(646.00,441.38)(57.509,-16.000){2}{\rule{0.600pt}{0.300pt}}
\multiput(706.00,425.13)(2.385,-0.490){24}{\rule{1.483pt}{0.118pt}}
\multiput(706.00,425.38)(57.923,-13.000){2}{\rule{0.741pt}{0.300pt}}
\multiput(767.00,412.13)(3.123,-0.486){18}{\rule{1.905pt}{0.117pt}}
\multiput(767.00,412.38)(57.046,-10.000){2}{\rule{0.953pt}{0.300pt}}
\multiput(828.00,402.14)(3.938,-0.482){14}{\rule{2.363pt}{0.116pt}}
\multiput(828.00,402.38)(56.097,-8.000){2}{\rule{1.181pt}{0.300pt}}
\multiput(889.00,394.14)(4.531,-0.480){12}{\rule{2.689pt}{0.116pt}}
\multiput(889.00,394.38)(55.418,-7.000){2}{\rule{1.345pt}{0.300pt}}
\multiput(950.00,387.14)(5.248,-0.475){10}{\rule{3.075pt}{0.115pt}}
\multiput(950.00,387.38)(53.618,-6.000){2}{\rule{1.538pt}{0.300pt}}
\multiput(1010.00,381.14)(6.499,-0.469){8}{\rule{3.735pt}{0.113pt}}
\multiput(1010.00,381.38)(53.248,-5.000){2}{\rule{1.868pt}{0.300pt}}
\multiput(1071.00,376.15)(8.339,-0.459){6}{\rule{4.650pt}{0.111pt}}
\multiput(1071.00,376.38)(51.349,-4.000){2}{\rule{2.325pt}{0.300pt}}
\multiput(1132.00,372.15)(8.339,-0.459){6}{\rule{4.650pt}{0.111pt}}
\multiput(1132.00,372.38)(51.349,-4.000){2}{\rule{2.325pt}{0.300pt}}
\multiput(1193.00,368.19)(22.706,-0.377){2}{\rule{9.225pt}{0.091pt}}
\multiput(1193.00,368.38)(41.853,-2.000){2}{\rule{4.613pt}{0.300pt}}
\multiput(1254.00,366.16)(11.594,-0.439){4}{\rule{6.075pt}{0.106pt}}
\multiput(1254.00,366.38)(47.391,-3.000){2}{\rule{3.038pt}{0.300pt}}
\multiput(1314.00,363.19)(22.706,-0.377){2}{\rule{9.225pt}{0.091pt}}
\multiput(1314.00,363.38)(41.853,-2.000){2}{\rule{4.613pt}{0.300pt}}
\multiput(1375.00,361.19)(22.706,-0.377){2}{\rule{9.225pt}{0.091pt}}
\multiput(1375.00,361.38)(41.853,-2.000){2}{\rule{4.613pt}{0.300pt}}
\sbox{\plotpoint}{\rule[-0.350pt]{0.700pt}{0.700pt}}%
\put(220,794){\usebox{\plotpoint}}
\multiput(221.20,789.61)(0.501,-0.631){116}{\rule{0.121pt}{1.059pt}}
\multiput(218.55,791.80)(61.000,-74.803){2}{\rule{0.700pt}{0.529pt}}
\multiput(282.20,712.99)(0.501,-0.565){116}{\rule{0.121pt}{0.967pt}}
\multiput(279.55,714.99)(61.000,-66.993){2}{\rule{0.700pt}{0.483pt}}
\multiput(342.00,646.30)(0.516,-0.501){110}{\rule{0.899pt}{0.121pt}}
\multiput(342.00,646.55)(58.134,-58.000){2}{\rule{0.450pt}{0.700pt}}
\multiput(402.00,588.30)(0.649,-0.501){88}{\rule{1.084pt}{0.121pt}}
\multiput(402.00,588.55)(58.751,-47.000){2}{\rule{0.542pt}{0.700pt}}
\multiput(463.00,541.30)(0.806,-0.501){70}{\rule{1.299pt}{0.121pt}}
\multiput(463.00,541.55)(58.305,-38.000){2}{\rule{0.649pt}{0.700pt}}
\multiput(524.00,503.30)(1.026,-0.502){54}{\rule{1.598pt}{0.121pt}}
\multiput(524.00,503.55)(57.683,-30.000){2}{\rule{0.799pt}{0.700pt}}
\multiput(585.00,473.30)(1.290,-0.502){42}{\rule{1.954pt}{0.121pt}}
\multiput(585.00,473.55)(56.944,-24.000){2}{\rule{0.977pt}{0.700pt}}
\multiput(646.00,449.30)(1.616,-0.503){32}{\rule{2.386pt}{0.121pt}}
\multiput(646.00,449.55)(55.049,-19.000){2}{\rule{1.193pt}{0.700pt}}
\multiput(706.00,430.30)(1.968,-0.504){26}{\rule{2.844pt}{0.121pt}}
\multiput(706.00,430.55)(55.098,-16.000){2}{\rule{1.422pt}{0.700pt}}
\multiput(767.00,414.29)(2.455,-0.505){20}{\rule{3.460pt}{0.122pt}}
\multiput(767.00,414.55)(53.819,-13.000){2}{\rule{1.730pt}{0.700pt}}
\multiput(828.00,401.29)(2.944,-0.506){16}{\rule{4.057pt}{0.122pt}}
\multiput(828.00,401.55)(52.580,-11.000){2}{\rule{2.028pt}{0.700pt}}
\multiput(889.00,390.29)(3.689,-0.508){12}{\rule{4.919pt}{0.122pt}}
\multiput(889.00,390.55)(50.789,-9.000){2}{\rule{2.460pt}{0.700pt}}
\multiput(950.00,381.29)(4.164,-0.509){10}{\rule{5.425pt}{0.123pt}}
\multiput(950.00,381.55)(48.740,-8.000){2}{\rule{2.712pt}{0.700pt}}
\multiput(1010.00,373.29)(6.146,-0.516){6}{\rule{7.292pt}{0.124pt}}
\multiput(1010.00,373.55)(45.866,-6.000){2}{\rule{3.646pt}{0.700pt}}
\multiput(1071.00,367.29)(6.146,-0.516){6}{\rule{7.292pt}{0.124pt}}
\multiput(1071.00,367.55)(45.866,-6.000){2}{\rule{3.646pt}{0.700pt}}
\multiput(1132.00,361.27)(15.961,-0.547){2}{\rule{10.850pt}{0.132pt}}
\multiput(1132.00,361.55)(38.480,-4.000){2}{\rule{5.425pt}{0.700pt}}
\multiput(1193.00,357.27)(15.961,-0.547){2}{\rule{10.850pt}{0.132pt}}
\multiput(1193.00,357.55)(38.480,-4.000){2}{\rule{5.425pt}{0.700pt}}
\put(1254,352.05){\rule{14.175pt}{0.700pt}}
\multiput(1254.00,353.55)(30.579,-3.000){2}{\rule{7.087pt}{0.700pt}}
\put(1314,349.05){\rule{14.408pt}{0.700pt}}
\multiput(1314.00,350.55)(31.095,-3.000){2}{\rule{7.204pt}{0.700pt}}
\put(1375,346.55){\rule{14.695pt}{0.700pt}}
\multiput(1375.00,347.55)(30.500,-2.000){2}{\rule{7.347pt}{0.700pt}}
\sbox{\plotpoint}{\rule[-0.400pt]{0.800pt}{0.800pt}}%
\put(220,542){\usebox{\plotpoint}}
\multiput(220,542)(22.636,-10.390){3}{\usebox{\plotpoint}}
\multiput(281,514)(22.636,-10.390){3}{\usebox{\plotpoint}}
\multiput(342,486)(22.570,-10.533){3}{\usebox{\plotpoint}}
\multiput(402,458)(22.912,-9.766){2}{\usebox{\plotpoint}}
\multiput(463,432)(23.305,-8.787){3}{\usebox{\plotpoint}}
\multiput(524,409)(23.667,-7.760){2}{\usebox{\plotpoint}}
\multiput(585,389)(23.888,-7.049){3}{\usebox{\plotpoint}}
\multiput(646,371)(24.163,-6.041){2}{\usebox{\plotpoint}}
\multiput(706,356)(24.360,-5.191){3}{\usebox{\plotpoint}}
\multiput(767,343)(24.511,-4.420){2}{\usebox{\plotpoint}}
\multiput(828,332)(24.640,-3.635){3}{\usebox{\plotpoint}}
\multiput(889,323)(24.695,-3.239){2}{\usebox{\plotpoint}}
\multiput(950,315)(24.783,-2.478){3}{\usebox{\plotpoint}}
\multiput(1010,309)(24.787,-2.438){2}{\usebox{\plotpoint}}
\multiput(1071,303)(24.853,-1.630){3}{\usebox{\plotpoint}}
\multiput(1132,299)(24.853,-1.630){2}{\usebox{\plotpoint}}
\multiput(1193,295)(24.877,-1.223){3}{\usebox{\plotpoint}}
\multiput(1254,292)(24.876,-1.244){2}{\usebox{\plotpoint}}
\multiput(1314,289)(24.893,-0.816){2}{\usebox{\plotpoint}}
\multiput(1375,287)(24.903,-0.408){3}{\usebox{\plotpoint}}
\put(1436,286){\usebox{\plotpoint}}
\put(220,181){\usebox{\plotpoint}}
\multiput(220,181)(24.579,4.029){3}{\usebox{\plotpoint}}
\multiput(281,191)(24.823,2.035){2}{\usebox{\plotpoint}}
\multiput(342,196)(24.893,0.830){3}{\usebox{\plotpoint}}
\multiput(402,198)(24.907,0.000){2}{\usebox{\plotpoint}}
\multiput(463,198)(24.903,-0.408){3}{\usebox{\plotpoint}}
\multiput(524,197)(24.893,-0.816){2}{\usebox{\plotpoint}}
\multiput(585,195)(24.893,-0.816){3}{\usebox{\plotpoint}}
\multiput(646,193)(24.876,-1.244){2}{\usebox{\plotpoint}}
\multiput(706,190)(24.893,-0.816){3}{\usebox{\plotpoint}}
\multiput(767,188)(24.903,-0.408){2}{\usebox{\plotpoint}}
\multiput(828,187)(24.903,-0.408){2}{\usebox{\plotpoint}}
\multiput(889,186)(24.903,-0.408){3}{\usebox{\plotpoint}}
\multiput(950,185)(24.903,-0.415){2}{\usebox{\plotpoint}}
\multiput(1010,184)(24.907,0.000){3}{\usebox{\plotpoint}}
\multiput(1071,184)(24.907,0.000){2}{\usebox{\plotpoint}}
\multiput(1132,184)(24.903,0.408){3}{\usebox{\plotpoint}}
\multiput(1193,185)(24.903,0.408){2}{\usebox{\plotpoint}}
\multiput(1254,186)(24.903,0.415){2}{\usebox{\plotpoint}}
\multiput(1314,187)(24.903,0.408){3}{\usebox{\plotpoint}}
\multiput(1375,188)(24.903,0.408){2}{\usebox{\plotpoint}}
\put(1436,189){\usebox{\plotpoint}}
\end{picture}

%% file: fig3.tex
\setlength{\unitlength}{0.240900pt}
\ifx\plotpoint\undefined\newsavebox{\plotpoint}\fi
\sbox{\plotpoint}{\rule[-0.150pt]{0.300pt}{0.300pt}}%
\begin{picture}(1500,900)(0,0)
\font\gnuplot=cmr10 at 10pt
\gnuplot
\sbox{\plotpoint}{\rule[-0.150pt]{0.300pt}{0.300pt}}%
\put(220.0,113.0){\rule[-0.150pt]{292.934pt}{0.300pt}}
\put(220.0,113.0){\rule[-0.150pt]{0.300pt}{184.048pt}}
\put(220.0,113.0){\rule[-0.150pt]{4.818pt}{0.300pt}}
\put(198,113){\makebox(0,0)[r]{$0$}}
\put(1416.0,113.0){\rule[-0.150pt]{4.818pt}{0.300pt}}
\put(220.0,266.0){\rule[-0.150pt]{4.818pt}{0.300pt}}
\put(198,266){\makebox(0,0)[r]{$0.2$}}
\put(1416.0,266.0){\rule[-0.150pt]{4.818pt}{0.300pt}}
\put(220.0,419.0){\rule[-0.150pt]{4.818pt}{0.300pt}}
\put(198,419){\makebox(0,0)[r]{$0.4$}}
\put(1416.0,419.0){\rule[-0.150pt]{4.818pt}{0.300pt}}
\put(220.0,571.0){\rule[-0.150pt]{4.818pt}{0.300pt}}
\put(198,571){\makebox(0,0)[r]{$0.6$}}
\put(1416.0,571.0){\rule[-0.150pt]{4.818pt}{0.300pt}}
\put(220.0,724.0){\rule[-0.150pt]{4.818pt}{0.300pt}}
\put(198,724){\makebox(0,0)[r]{$0.8$}}
\put(1416.0,724.0){\rule[-0.150pt]{4.818pt}{0.300pt}}
\put(220.0,877.0){\rule[-0.150pt]{4.818pt}{0.300pt}}
\put(198,877){\makebox(0,0)[r]{$1$}}
\put(1416.0,877.0){\rule[-0.150pt]{4.818pt}{0.300pt}}
\put(220.0,113.0){\rule[-0.150pt]{0.300pt}{4.818pt}}
\put(220,68){\makebox(0,0){$0$}}
\put(220.0,857.0){\rule[-0.150pt]{0.300pt}{4.818pt}}
\put(423.0,113.0){\rule[-0.150pt]{0.300pt}{4.818pt}}
\put(423,68){\makebox(0,0){$0.05$}}
\put(423.0,857.0){\rule[-0.150pt]{0.300pt}{4.818pt}}
\put(625.0,113.0){\rule[-0.150pt]{0.300pt}{4.818pt}}
\put(625,68){\makebox(0,0){$0.1$}}
\put(625.0,857.0){\rule[-0.150pt]{0.300pt}{4.818pt}}
\put(828.0,113.0){\rule[-0.150pt]{0.300pt}{4.818pt}}
\put(828,68){\makebox(0,0){$0.15$}}
\put(828.0,857.0){\rule[-0.150pt]{0.300pt}{4.818pt}}
\put(1031.0,113.0){\rule[-0.150pt]{0.300pt}{4.818pt}}
\put(1031,68){\makebox(0,0){$0.2$}}
\put(1031.0,857.0){\rule[-0.150pt]{0.300pt}{4.818pt}}
\put(1233.0,113.0){\rule[-0.150pt]{0.300pt}{4.818pt}}
\put(1233,68){\makebox(0,0){$0.25$}}
\put(1233.0,857.0){\rule[-0.150pt]{0.300pt}{4.818pt}}
\put(1436.0,113.0){\rule[-0.150pt]{0.300pt}{4.818pt}}
\put(1436,68){\makebox(0,0){$0.3$}}
\put(1436.0,857.0){\rule[-0.150pt]{0.300pt}{4.818pt}}
\put(220.0,113.0){\rule[-0.150pt]{292.934pt}{0.300pt}}
\put(1436.0,113.0){\rule[-0.150pt]{0.300pt}{184.048pt}}
\put(220.0,877.0){\rule[-0.150pt]{292.934pt}{0.300pt}}
\put(45,495){\makebox(0,0){$r$}}
\put(828,23){\makebox(0,0){$s$}}
\put(605,686){\makebox(0,0)[l]{$1-\rho(0)=0$}}
\put(504,304){\makebox(0,0)[l]{$1-\rho(0)=0.01$}}
\put(1031,571){\makebox(0,0)[l]{$1-\rho(0)=-0.01$}}
\put(220.0,113.0){\rule[-0.150pt]{0.300pt}{184.048pt}}
\multiput(224.00,876.13)(0.807,-0.494){44}{\rule{0.558pt}{0.119pt}}
\multiput(224.00,876.38)(35.843,-23.000){2}{\rule{0.279pt}{0.300pt}}
\multiput(261.00,853.13)(0.957,-0.494){40}{\rule{0.646pt}{0.119pt}}
\multiput(261.00,853.38)(38.658,-21.000){2}{\rule{0.323pt}{0.300pt}}
\multiput(301.00,832.13)(0.967,-0.497){82}{\rule{0.654pt}{0.120pt}}
\multiput(301.00,832.38)(79.643,-42.000){2}{\rule{0.327pt}{0.300pt}}
\multiput(382.00,790.13)(0.936,-0.494){42}{\rule{0.634pt}{0.119pt}}
\multiput(382.00,790.38)(39.684,-22.000){2}{\rule{0.317pt}{0.300pt}}
\multiput(423.00,768.13)(0.873,-0.494){44}{\rule{0.597pt}{0.119pt}}
\multiput(423.00,768.38)(38.761,-23.000){2}{\rule{0.298pt}{0.300pt}}
\multiput(463.00,745.13)(0.936,-0.494){42}{\rule{0.634pt}{0.119pt}}
\multiput(463.00,745.38)(39.684,-22.000){2}{\rule{0.317pt}{0.300pt}}
\multiput(504.00,723.13)(0.873,-0.494){44}{\rule{0.597pt}{0.119pt}}
\multiput(504.00,723.38)(38.761,-23.000){2}{\rule{0.298pt}{0.300pt}}
\multiput(544.00,700.13)(0.882,-0.497){90}{\rule{0.603pt}{0.120pt}}
\multiput(544.00,700.38)(79.748,-46.000){2}{\rule{0.302pt}{0.300pt}}
\multiput(625.00,654.13)(0.845,-0.497){94}{\rule{0.581pt}{0.120pt}}
\multiput(625.00,654.38)(79.794,-48.000){2}{\rule{0.291pt}{0.300pt}}
\multiput(706.00,606.13)(0.822,-0.495){48}{\rule{0.567pt}{0.119pt}}
\multiput(706.00,606.38)(39.823,-25.000){2}{\rule{0.284pt}{0.300pt}}
\multiput(747.00,581.13)(0.802,-0.495){48}{\rule{0.555pt}{0.119pt}}
\multiput(747.00,581.38)(38.848,-25.000){2}{\rule{0.278pt}{0.300pt}}
\multiput(787.00,556.13)(0.822,-0.495){48}{\rule{0.567pt}{0.119pt}}
\multiput(787.00,556.38)(39.823,-25.000){2}{\rule{0.284pt}{0.300pt}}
\multiput(828.00,531.13)(0.761,-0.495){52}{\rule{0.531pt}{0.119pt}}
\multiput(828.00,531.38)(39.899,-27.000){2}{\rule{0.265pt}{0.300pt}}
\multiput(869.00,504.13)(0.742,-0.495){52}{\rule{0.519pt}{0.119pt}}
\multiput(869.00,504.38)(38.922,-27.000){2}{\rule{0.260pt}{0.300pt}}
\multiput(909.00,477.13)(0.708,-0.496){56}{\rule{0.499pt}{0.119pt}}
\multiput(909.00,477.38)(39.964,-29.000){2}{\rule{0.250pt}{0.300pt}}
\multiput(950.00,448.13)(0.664,-0.498){120}{\rule{0.473pt}{0.120pt}}
\multiput(950.00,448.38)(80.018,-61.000){2}{\rule{0.237pt}{0.300pt}}
\multiput(1031.00,387.13)(0.588,-0.496){66}{\rule{0.428pt}{0.120pt}}
\multiput(1031.00,387.38)(39.112,-34.000){2}{\rule{0.214pt}{0.300pt}}
\multiput(1071.00,353.13)(0.554,-0.497){72}{\rule{0.407pt}{0.120pt}}
\multiput(1071.00,353.38)(40.154,-37.000){2}{\rule{0.204pt}{0.300pt}}
\multiput(1112.37,315.40)(0.496,-0.516){56}{\rule{0.119pt}{0.385pt}}
\multiput(1111.38,316.20)(29.000,-29.200){2}{\rule{0.300pt}{0.193pt}}
\multiput(1141.38,285.22)(0.488,-0.590){20}{\rule{0.117pt}{0.430pt}}
\multiput(1140.38,286.11)(11.000,-12.108){2}{\rule{0.300pt}{0.215pt}}
\multiput(1152.38,272.21)(0.494,-0.595){40}{\rule{0.119pt}{0.432pt}}
\multiput(1151.38,273.10)(21.000,-24.103){2}{\rule{0.300pt}{0.216pt}}
\multiput(1173.38,246.88)(0.494,-0.727){38}{\rule{0.119pt}{0.510pt}}
\multiput(1172.38,247.94)(20.000,-27.941){2}{\rule{0.300pt}{0.255pt}}
\multiput(1193.38,217.45)(0.494,-0.904){38}{\rule{0.119pt}{0.615pt}}
\multiput(1192.38,218.72)(20.000,-34.724){2}{\rule{0.300pt}{0.308pt}}
\multiput(1213.38,180.00)(0.495,-1.489){46}{\rule{0.119pt}{0.963pt}}
\multiput(1212.38,182.00)(24.000,-69.002){2}{\rule{0.300pt}{0.481pt}}
\put(229,123){\usebox{\plotpoint}}
\multiput(229.40,123.00)(0.439,1.456){4}{\rule{0.106pt}{0.875pt}}
\multiput(228.38,123.00)(3.000,6.184){2}{\rule{0.300pt}{0.438pt}}
\multiput(232.38,131.00)(0.482,2.184){14}{\rule{0.116pt}{1.350pt}}
\multiput(231.38,131.00)(8.000,31.198){2}{\rule{0.300pt}{0.675pt}}
\multiput(240.38,165.00)(0.494,1.825){40}{\rule{0.119pt}{1.161pt}}
\multiput(239.38,165.00)(21.000,73.591){2}{\rule{0.300pt}{0.580pt}}
\multiput(261.37,241.00)(0.497,1.216){78}{\rule{0.120pt}{0.803pt}}
\multiput(260.38,241.00)(40.000,95.334){2}{\rule{0.300pt}{0.401pt}}
\multiput(301.37,338.00)(0.498,0.568){160}{\rule{0.120pt}{0.416pt}}
\multiput(300.38,338.00)(81.000,91.137){2}{\rule{0.300pt}{0.208pt}}
\multiput(382.00,430.37)(1.098,0.497){72}{\rule{0.732pt}{0.120pt}}
\multiput(382.00,429.38)(79.481,37.000){2}{\rule{0.366pt}{0.300pt}}
\multiput(463.00,467.38)(2.334,0.485){16}{\rule{1.442pt}{0.117pt}}
\multiput(463.00,466.38)(38.008,9.000){2}{\rule{0.721pt}{0.300pt}}
\multiput(504.00,476.43)(14.782,0.377){2}{\rule{6.075pt}{0.091pt}}
\multiput(504.00,475.38)(27.391,2.000){2}{\rule{3.038pt}{0.300pt}}
\put(544,476.88){\rule{9.877pt}{0.300pt}}
\multiput(544.00,477.38)(20.500,-1.000){2}{\rule{4.938pt}{0.300pt}}
\multiput(585.00,476.14)(4.247,-0.469){8}{\rule{2.475pt}{0.113pt}}
\multiput(585.00,476.38)(34.863,-5.000){2}{\rule{1.238pt}{0.300pt}}
\multiput(625.00,471.14)(3.036,-0.480){12}{\rule{1.832pt}{0.116pt}}
\multiput(625.00,471.38)(37.197,-7.000){2}{\rule{0.916pt}{0.300pt}}
\multiput(666.00,464.13)(2.042,-0.486){18}{\rule{1.275pt}{0.117pt}}
\multiput(666.00,464.38)(37.354,-10.000){2}{\rule{0.638pt}{0.300pt}}
\multiput(706.00,454.13)(1.735,-0.489){22}{\rule{1.100pt}{0.118pt}}
\multiput(706.00,454.38)(38.717,-12.000){2}{\rule{0.550pt}{0.300pt}}
\multiput(747.00,442.13)(1.445,-0.491){26}{\rule{0.932pt}{0.118pt}}
\multiput(747.00,442.38)(38.065,-14.000){2}{\rule{0.466pt}{0.300pt}}
\multiput(787.00,428.13)(1.293,-0.492){30}{\rule{0.844pt}{0.118pt}}
\multiput(787.00,428.38)(39.249,-16.000){2}{\rule{0.422pt}{0.300pt}}
\multiput(828.00,412.13)(1.216,-0.492){32}{\rule{0.799pt}{0.119pt}}
\multiput(828.00,412.38)(39.343,-17.000){2}{\rule{0.399pt}{0.300pt}}
\multiput(869.00,395.13)(1.005,-0.494){38}{\rule{0.675pt}{0.119pt}}
\multiput(869.00,395.38)(38.599,-20.000){2}{\rule{0.338pt}{0.300pt}}
\multiput(909.00,375.13)(0.936,-0.494){42}{\rule{0.634pt}{0.119pt}}
\multiput(909.00,375.38)(39.684,-22.000){2}{\rule{0.317pt}{0.300pt}}
\multiput(950.00,353.13)(0.802,-0.495){48}{\rule{0.555pt}{0.119pt}}
\multiput(950.00,353.38)(38.848,-25.000){2}{\rule{0.278pt}{0.300pt}}
\multiput(990.00,328.13)(0.761,-0.495){52}{\rule{0.531pt}{0.119pt}}
\multiput(990.00,328.38)(39.899,-27.000){2}{\rule{0.265pt}{0.300pt}}
\multiput(1031.00,301.13)(0.645,-0.496){60}{\rule{0.462pt}{0.119pt}}
\multiput(1031.00,301.38)(39.041,-31.000){2}{\rule{0.231pt}{0.300pt}}
\multiput(1071.00,270.13)(0.539,-0.497){74}{\rule{0.399pt}{0.120pt}}
\multiput(1071.00,270.38)(40.173,-38.000){2}{\rule{0.199pt}{0.300pt}}
\multiput(1112.37,231.19)(0.497,-0.600){78}{\rule{0.120pt}{0.435pt}}
\multiput(1111.38,232.10)(40.000,-47.097){2}{\rule{0.300pt}{0.218pt}}
\multiput(1152.38,182.82)(0.482,-0.755){14}{\rule{0.116pt}{0.525pt}}
\multiput(1151.38,183.91)(8.000,-10.910){2}{\rule{0.300pt}{0.263pt}}
\multiput(1160.38,170.51)(0.482,-0.885){14}{\rule{0.116pt}{0.600pt}}
\multiput(1159.38,171.75)(8.000,-12.755){2}{\rule{0.300pt}{0.300pt}}
\multiput(1168.38,156.61)(0.485,-0.841){16}{\rule{0.117pt}{0.575pt}}
\multiput(1167.38,157.81)(9.000,-13.807){2}{\rule{0.300pt}{0.288pt}}
\multiput(1177.38,140.73)(0.482,-1.210){14}{\rule{0.116pt}{0.788pt}}
\multiput(1176.38,142.37)(8.000,-17.366){2}{\rule{0.300pt}{0.394pt}}
\multiput(1185.43,121.58)(0.377,-1.575){2}{\rule{0.091pt}{0.825pt}}
\multiput(1184.38,123.29)(2.000,-3.288){2}{\rule{0.300pt}{0.413pt}}
\put(798,877){\usebox{\plotpoint}}
\multiput(798.38,873.58)(0.475,-1.286){10}{\rule{0.115pt}{0.825pt}}
\multiput(797.38,875.29)(6.000,-13.288){2}{\rule{0.300pt}{0.413pt}}
\multiput(804.39,858.89)(0.459,-1.177){6}{\rule{0.111pt}{0.750pt}}
\multiput(803.38,860.44)(4.000,-7.443){2}{\rule{0.300pt}{0.375pt}}
\multiput(808.39,849.58)(0.459,-1.315){6}{\rule{0.111pt}{0.825pt}}
\multiput(807.38,851.29)(4.000,-8.288){2}{\rule{0.300pt}{0.413pt}}
\multiput(812.39,839.89)(0.459,-1.177){6}{\rule{0.111pt}{0.750pt}}
\multiput(811.38,841.44)(4.000,-7.443){2}{\rule{0.300pt}{0.375pt}}
\multiput(816.39,831.20)(0.459,-1.040){6}{\rule{0.111pt}{0.675pt}}
\multiput(815.38,832.60)(4.000,-6.599){2}{\rule{0.300pt}{0.338pt}}
\multiput(820.39,822.89)(0.459,-1.177){6}{\rule{0.111pt}{0.750pt}}
\multiput(819.38,824.44)(4.000,-7.443){2}{\rule{0.300pt}{0.375pt}}
\multiput(824.39,814.20)(0.459,-1.040){6}{\rule{0.111pt}{0.675pt}}
\multiput(823.38,815.60)(4.000,-6.599){2}{\rule{0.300pt}{0.338pt}}
\multiput(828.37,806.47)(0.497,-0.892){80}{\rule{0.120pt}{0.609pt}}
\multiput(827.38,807.74)(41.000,-71.736){2}{\rule{0.300pt}{0.305pt}}
\multiput(869.37,733.79)(0.497,-0.764){78}{\rule{0.120pt}{0.533pt}}
\multiput(868.38,734.89)(40.000,-59.895){2}{\rule{0.300pt}{0.266pt}}
\multiput(909.37,673.02)(0.497,-0.671){80}{\rule{0.120pt}{0.477pt}}
\multiput(908.38,674.01)(41.000,-54.009){2}{\rule{0.300pt}{0.239pt}}
\multiput(950.37,618.13)(0.497,-0.625){78}{\rule{0.120pt}{0.450pt}}
\multiput(949.38,619.07)(40.000,-49.066){2}{\rule{0.300pt}{0.225pt}}
\multiput(990.37,568.23)(0.497,-0.585){80}{\rule{0.120pt}{0.426pt}}
\multiput(989.38,569.12)(41.000,-47.115){2}{\rule{0.300pt}{0.213pt}}
\multiput(1031.37,520.26)(0.497,-0.575){78}{\rule{0.120pt}{0.420pt}}
\multiput(1030.38,521.13)(40.000,-45.128){2}{\rule{0.300pt}{0.210pt}}
\multiput(1071.37,474.26)(0.497,-0.573){80}{\rule{0.120pt}{0.419pt}}
\multiput(1070.38,475.13)(41.000,-46.131){2}{\rule{0.300pt}{0.209pt}}
\multiput(1112.37,427.19)(0.497,-0.600){78}{\rule{0.120pt}{0.435pt}}
\multiput(1111.38,428.10)(40.000,-47.097){2}{\rule{0.300pt}{0.218pt}}
\multiput(1152.37,379.11)(0.497,-0.634){80}{\rule{0.120pt}{0.455pt}}
\multiput(1151.38,380.05)(41.000,-51.055){2}{\rule{0.300pt}{0.228pt}}
\multiput(1193.37,326.76)(0.497,-0.776){78}{\rule{0.120pt}{0.540pt}}
\multiput(1192.38,327.88)(40.000,-60.879){2}{\rule{0.300pt}{0.270pt}}
\multiput(1233.38,264.38)(0.494,-0.933){40}{\rule{0.119pt}{0.632pt}}
\multiput(1232.38,265.69)(21.000,-37.688){2}{\rule{0.300pt}{0.316pt}}
\multiput(1254.38,224.26)(0.494,-1.386){38}{\rule{0.119pt}{0.900pt}}
\multiput(1253.38,226.13)(20.000,-53.132){2}{\rule{0.300pt}{0.450pt}}
\multiput(1274.38,166.31)(0.482,-2.639){14}{\rule{0.116pt}{1.613pt}}
\multiput(1273.38,169.65)(8.000,-37.653){2}{\rule{0.300pt}{0.806pt}}
\put(1281.88,124){\rule{0.300pt}{1.927pt}}
\multiput(1281.38,128.00)(1.000,-4.000){2}{\rule{0.300pt}{0.964pt}}
\put(1282.88,114){\rule{0.300pt}{1.445pt}}
\multiput(1282.38,117.00)(1.000,-3.000){2}{\rule{0.300pt}{0.723pt}}
\put(1283.0,120.0){\rule[-0.150pt]{0.300pt}{0.964pt}}
\end{picture}